# Anisotropic intrinsic spin Hall effect in quantum wires

A.W. Cummings, R. Akis, and D.K. Ferry

Department of Electrical Engineering and Center for Solid State Electronics Research,
Arizona State University, Tempe, Arizona 85287-5706

We use numerical simulations to investigate the spin Hall effect in quantum wires in the presence of both Rashba and Dresselhaus spin-orbit coupling. We find that the intrinsic spin Hall effect is highly anisotropic with respect to the orientation of the wire, and that the nature of this anisotropy depends strongly on the electron density and the relative strengths of the Rashba and Dresselhaus spin-orbit coupling. In particular, at low densities when only one subband of the quantum wire is occupied, the spin Hall effect is strongest for electron momentum along the $[\bar{1}10]$ axis, which is opposite than what is expected for the purely 2D case. In addition, when more than one subband is occupied, the strength and anisotropy of the spin Hall effect can vary greatly over relatively small changes in electron density, which makes it difficult to predict which wire orientation will maximize the strength of the spin Hall effect. These results help to illuminate the role of quantum confinement in spin-orbit-coupled systems, and can serve as a guide for future experimental work on the use of quantum wires for spin-Hall-based spintronic applications.

In recent years, much effort has been spent on the study of spintronics as an alternative method of information processing and storage. Spintronics is a field of electronics that, instead of using an electron's charge, utilizes its spin as the basic unit of information [1]. The spin of an electron can be manipulated in a variety of ways, but in order to take advantage of current semiconductor processing technology, it would be ideal to find a purely electrical means of doing so. For this reason, a great deal of attention has centered on the spin Hall effect in semiconductors, where in the presence of spin-orbit coupling a transverse spin current arises in response to a longitudinal charge current, without the need for magnetic materials or externally applied magnetic fields [2-4]. Spin-orbit coupling also plays a fundamental role in the emerging field of topological insulators, which are characterized by an insulating bulk material with conducting edge states. These edge states are topologically protected from disorder, and show potential for both spintronic and quantum computing applications [5,6]. Therefore, an examination of the interplay between the different types of spin-orbit coupling in semiconductors is important to further the understanding of both of these fields of study.

The spin Hall effect most commonly originates from the Rashba form of spin-orbit coupling [7], which is present in a two-dimensional electron gas (2DEG) formed in an asymmetric semiconductor quantum well. Early studies showed that in the infinite 2D limit, arbitrarily small disorder introduces a vertex correction that exactly cancels out the transverse spin current [8]. However, in finite systems such as quantum wires, the spin Hall effect survives in the presence of disorder [9] and manifests itself as an accumulation of oppositely polarized spins on opposite sides of the wire [10-12]. This has led to the proposal of a variety of devices that utilize branched, quasi-1D structures to generate and detect spin-polarized currents through purely electrical measurements [13-18].

In addition to Rashba spin-orbit coupling, a term due to the inversion asymmetry of the host semiconductor crystal, known as Dresselhaus spin-orbit coupling [19], can also yield a spin Hall current. When both forms of spin-orbit coupling are present in a two-dimensional system, they interfere with one another, and this interference is anisotropic with respect to the direction of the charge carrier's momentum [20]. Several studies have investigated the nature of the spin Hall effect in the presence of both Rashba and Dresselhaus spin-orbit coupling. Early on, it was shown that in a 2DEG the spin Hall conductivity can change sign when the strength of the Rashba coupling becomes smaller than the Dresselhaus coupling [21,22]. This effect was also calculated in a quasi-1D system in the presence of disorder [23]. In the presence of an in-plane magnetic field, it was shown that the spin Hall conductance in a 2DEG has an anisotropic dependence on the direction of the field [24,25], while Trushim *et al.* showed that the spin accumulation due to the *extrinsic* spin Hall effect has a similar dependence on the direction of electron momentum as that shown in Fig. 1a below [26]. Finally, Wang *et al.* highlighted the anisotropy of both the charge conductance and the spin precession rate in quantum waveguides [27]. However, none of these works have examined the anisotropic nature of the intrinsic spin Hall effect in quasi-1D systems.

In this paper, we present numerical calculations of the strength of the intrinsic spin Hall effect in quantum wires as a function of the orientation of the wire with respect to the underlying crystal structure. We find that the intrinsic spin Hall effect is highly anisotropic, and that the nature of this anisotropy depends strongly on the electron density and the relative strengths of the Rashba and Dresselhaus spin-orbit coupling. These results differ significantly from the pure 2D case and can help to guide experimental work on the use of quantum wires in spin-Hall-based spintronic applications.

To examine the combined effect of Rashba and Dresselhaus spin-orbit coupling, we start with the two-band model of the conduction band in an asymmetric [001]-grown quantum well, in a III-V semiconductor such as GaAs or InAs. The Hamiltonian of this system is given by [20]

$$H = \frac{\hbar^2 k^2}{2m^*} + V(x,y) + \alpha(k_y \sigma_x - k_x \sigma_y) \\ + \beta(k_y \sigma_y - k_x \sigma_x) + \eta k_x k_y (k_y \sigma_x - k_x \sigma_y),$$

(1)

where the first two terms are the usual kinetic and potential energies, and the third term is the Rashba spin-orbit coupling due to the asymmetry of the quantum well. The parameter $\alpha$ characterizes the strength of the Rashba coupling and can be tuned with the application of a top-gate or back-gate voltage [28,29]. The last two terms form the Dresselhaus spin-orbit coupling that arises from the bulk inversion asymmetry of the semiconductor crystal. In a quantum well this is broken into a linear term and a cubic term, where the linear term arises from the confinement of the momentum operator along the $z$-axis. The parameter $\beta$ characterizes the strength of the linear portion of the Dresselhaus interaction, and is given by $\eta \langle k_z^2 \rangle$, where $\eta$ is a material-dependent parameter [20]. In this work, the cubic part of the Dresselhaus spin-orbit coupling is ignored, as it was found to have a negligible effect on the results. The variables $k_x$ and $k_y$ refer to the electron momentum along the [100] and [010] crystal axes, respectively, and the $\sigma_i$ are the Pauli matrices representing the spin of the electron.

The Rashba and the linear Dresselhaus terms in (1) have similar behavior in that they both result in a *k*-linear energy splitting of the conduction band in an infinite 2DEG. When only one of these two terms is present, the splitting is independent of the direction of the electron momentum *k*, and is given by

$$E_{\pm}(k) = \frac{\hbar^2 k^2}{2m^*} \pm k\alpha \quad \text{or} \quad E_{\pm}(k) = \frac{\hbar^2 k^2}{2m^*} \pm k\beta. \quad (2)$$

However, when both terms are present, the splitting of the conduction band takes on a more complicated form,

$$E_{\pm}(k) = \frac{\hbar^2 k^2}{2m^*} \pm k\sqrt{\alpha^2 + \beta^2 - 2\alpha\beta \sin(2\theta)}, \quad (3)$$

where $\theta$ is the angle of *k* with respect to the [100] crystal axis [20]. As seen in (3), the magnitude of the splitting is still linear in *k*, but also shows anisotropic behavior. In particular, the splitting is minimized when the electron momentum is along the [110] axis ($\theta = 45°$), and is maximized along the [$\bar{1}$10] axis ($\theta = 135°$).

The Rashba and linear Dresselhaus terms also each yield a spin Hall current, perpendicular to the electron momentum and polarized out of the plane of the 2DEG [3]. The spin Hall currents due to each of these mechanisms are opposite to one another, and when only one is present, the spin current magnitude is independent of the direction of *k*. However, when both terms are present, the magnitude of the spin current becomes anisotropic in a manner similar to the energy splitting of the conduction band,

$$\left| \langle \mathbf{j}_s^z \rangle_{\pm} \right| = \frac{1}{2}\sqrt{\alpha^2 + \beta^2 + 2\alpha\beta \sin(2\phi)}, \quad (4)$$

where $\tan\phi = \dfrac{\alpha\sin\theta - \beta\cos\theta}{\alpha\cos\theta - \beta\sin\theta}$. In (4), $\langle \mathbf{j}_s^z \rangle_{\pm}$ is the expectation value of the basic spin current operator for each subband, and is calculated as $\langle \mathbf{j}_s^z \rangle_\pm = \hbar/2 \cdot \langle \pm | \sigma_z \mathbf{v} | \pm \rangle$, where $|\pm\rangle = \frac{1}{\sqrt{2}} [e^{-i\phi} \quad \pm i]^T$ are the eigenstates corresponding to $E_\pm$ in (3). It should be noted that $\langle \mathbf{j}_s^z \rangle_\pm$ does not contain any contributions along the direction of the electron momentum and thus represents the pure spin Hall current for each subband. The behavior of (4) is shown in Fig. 1a, where $\left| \langle \mathbf{j}_s^z \rangle_\pm \right|$ is plotted as a function of $\theta$ for $\beta = 5$ meV-nm and $\alpha = 5.1$ (bottom curve), 6, 7, 8, …, 20 meV-nm (top curve). In this figure, one can see that when both Rashba and Dresselhaus spin-orbit coupling are present, the magnitude of the spin Hall current in the 2DEG is maximized for electron momentum along the [110] axis and is minimized along the [$\bar{1}$10] axis. When $\alpha \gg \beta$, the dependence on $\theta$ is nearly sinusoidal. However, for $\alpha \approx \beta$, the magnitude of the spin Hall current shows a sharp peak along [110] and is nearly zero everywhere else. The energy splitting given by (3) is shown in Fig. 1b, where the bottom curve is for $\alpha = 5.1$ meV-nm, and the top curve is for $\alpha = 20$ meV-nm. A comparison of Figs. 1a and 1b indicates a maximum in the spin Hall current at those points where the energy splitting is the smallest. In the topological picture, the minima in the energy splitting represent anticrossing points between the bands in momentum space. These anticrossing points correspond to an enhancement of the Berry curvature of each band, which leads to an amplification of the spin Hall current of each band [30,31].

While Fig. 1 provides a picture of the anisotropy of the spin Hall effect in 2D systems, it does not consider the lateral confinement that exists in a quantum wire. It has been shown that this confinement plays a key role in the strength of the intrinsic spin Hall effect in quasi-1D systems [31], and we therefore turn to numerical calculations which take this into account. As a basis for our numerical calculations, we assume a quantum wire formed in a 2DEG, with a Hamiltonian given by (1). We assume a smooth transition from the 2DEG to the quasi-1D region, such that resonant states along the length of the wire can be ignored. In addition, we assume that the electron density in the 2DEG sets the Fermi energy in the quasi-1D region. To allow for the longitudinal axis of the wire to lie along an arbitrary crystal direction, we make the transformation

$$k_x = k_\parallel \cos\theta + k_\perp \sin\theta \text{ and } k_y = k_\parallel \sin\theta + k_\perp \cos\theta, \tag{5}$$

where $k_\parallel$ represents the electron momentum along the length of the wire and $k_\perp$ is the momentum along its transverse axis. Similarly, $x_\parallel$ and $x_\perp$ represent the spatial coordinates along these axes. We assume hard wall boundaries by defining the confining potential, $V(x_\perp, x_\parallel)$, to be 0 when $0 < x_\perp < W$ and infinite otherwise, where $W$ is the width of the wire.

Because the Hamiltonian is translationally invariant along the longitudinal axis of the wire, the electron wave function can be written as

$$\psi(x_\perp, x_\parallel) = e^{ik_\parallel x_\parallel} \begin{bmatrix} \phi_\uparrow(x_\perp) \\ \phi_\downarrow(x_\perp) \end{bmatrix}, \tag{6}$$

where $\phi_\uparrow$ and $\phi_\downarrow$ are the spin components of the wave function oriented out of the plane of the 2DEG along +z and –z, respectively. The wave numbers $k_\perp$ and $k_\parallel$ can be written in their operator form as $k_\mu = -i\partial/\partial x_\mu$, and (1), (5), and (6) can be substituted into the time-independent Schrödinger equation, $H\psi = E\psi$. After some algebra and discretizing along the $x_\perp$-axis, the Schrödinger equation becomes

$$(\mathbf{H}_0 + \mathbf{H}_1)\phi = E\phi, \tag{7}$$

where $\phi = [\phi_\uparrow(x_\perp) \ \phi_\downarrow(x_\perp)]^T$. $\mathbf{H}_0$ represents the bare kinetic energy term in (1), and $\mathbf{H}_1$ includes the contributions of the Rashba spin-orbit coupling and the linear part of the Dresselhaus spin-orbit coupling. The cubic part of the Dresselhaus term was found to have a negligible effect on the final results and thus for simplicity is left out of (7). The matrix $\mathbf{H}_0$ is given by

$$\mathbf{H}_0 = \begin{bmatrix} \mathbf{H}_0^\uparrow & \mathbf{0} \\ \mathbf{0} & \mathbf{H}_0^\downarrow \end{bmatrix}, \tag{8a}$$

$$\mathbf{H}_0^{\uparrow,\downarrow} = \begin{bmatrix} h_{11}^{(0)} & -t & & \mathbf{0} \\ -t & h_{22}^{(0)} & \ddots & \\ & \ddots & \ddots & -t \\ \mathbf{0} & & -t & h_{NN}^{(0)} \end{bmatrix}, \tag{8b}$$

where $h_{nn}^{(0)} = 2t + \hbar^2 k_\parallel^2 / 2m^*$, $t = \hbar^2 / 2m^* a^2$, $a$ is the grid spacing along the $x_\perp$-axis, and $N$ is the number of grid points along the $x_\perp$-axis. The matrix $\mathbf{H}_1$ is given by

$$\mathbf{H}_1 = \begin{bmatrix} \mathbf{0} & \mathbf{H}_1^\downarrow \\ \mathbf{H}_1^\uparrow & \mathbf{0} \end{bmatrix}, \tag{9a}$$

$$\mathbf{H}_1^\uparrow = \left(\mathbf{H}_1^\downarrow\right)^\dagger = \begin{bmatrix} h_{11}^{(1)} & h_{12}^{(1)} & & \mathbf{0} \\ h_{21}^{(1)} & h_{22}^{(1)} & \ddots & \\ & \ddots & \ddots & h_{N-1,N}^{(1)} \\ \mathbf{0} & & h_{N,N-1}^{(1)} & h_{NN}^{(1)} \end{bmatrix}, \tag{9b}$$

where $h_{nn}^{(1)} = -i\chi_{SO}^- k_\parallel$, $h_{n,n\pm1}^{(1)} = \pm i\chi_{SO}^+/2a$, and $\chi_{SO}^\pm = \alpha e^{i\theta} \pm i\beta e^{-i\theta}$.

With the matrices given in (8) and (9), (7) can be used to find the energies and the corresponding wave functions of a quasi-1D wire as a function of $k_\parallel$ for given values of the Rashba spin-orbit coupling, the Dresselhaus spin-orbit coupling, the wire width, the effective mass, and the angle the wire makes with respect to the [100] crystal axis. Once the transverse wave functions are known, the strength of the spin Hall effect can be calculated. As in Ref. 28, we characterize the strength of the spin Hall effect in a particular subband with the spin displacement operator,

$$\Delta x_s = \langle x_\perp \sigma_z \rangle = \int_{x_\perp} \phi_\uparrow^* x_\perp \phi_\uparrow dx_\perp - \int_{x_\perp} \phi_\downarrow^* x_\perp \phi_\downarrow dx_\perp = \langle x_\perp^\uparrow \rangle - \langle x_\perp^\downarrow \rangle. \qquad (10)$$

The spin displacement operator gives the transverse separation of the spin-up and spin-down wave functions and allows a direct comparison of the strength of the spin Hall effect to the width of the quantum wire. Furthermore, the spin displacement has been shown to be proportional to the efficiency of spin filters based on branching structures in quantum wires [32]. If we assume low temperatures and low bias voltages, then electron transport occurs only at energies near the Fermi energy. Thus, the spin displacement of the current-carrying modes in the wire is calculated at the points where the subbands cross the Fermi energy with a positive slope, *dE/dk*.

In our simulations, we considered a 100-nm-wide GaAs quantum wire with $\alpha = 1$, 4.9, 5.1, 6, 10, and 20 meV-nm, and $\beta = 5$ meV-nm. Figure 2 shows the magnitude of $\Delta x_s$ as a function of the electron density and the orientation of the wire, for each value of $\alpha$. The orientation of the wire was varied from 0° to 360° with respect to the [100] crystal axis, and the electron density was varied from zero up to 2.8 x $10^{11}$ cm$^{-2}$. In this figure, a couple of primary trends can be identified. The first is that as the magnitude of $\alpha$ decreases, the overall magnitude of $\Delta x_s$ also decreases, from a peak of 12.3 nm in Fig. 2a to an average close to zero in Figs. 2d and 2e. Because the Rashba and Dresselhaus spin-orbit interactions yield opposite spin currents, it is understandable that the overall magnitude of $\Delta x_s$ would be smallest when $\alpha \approx \beta$, as seen in Figs. 2d and 2e. These general results correspond to what is seen in Fig. 1a in the 2D limit. When $\alpha < \beta$, as is the case in Figs. 2e and 2f, then the direction of the spin displacement reverses, echoing the results of Refs. 18-20. The second primary trend is that as the electron density increases and more subbands are populated, the average value of $\Delta x_s$ also decreases, due to interference between the subbands. In Fig. 2, the onset of each subband is distinguished by the horizontal transitions at densities of 0.6, 1.3, and 2.4 x $10^{11}$ cm$^{-2}$.

While the trends described above provide some high-level information about the nature of the spin Hall effect in quantum wires, the most interesting feature of Fig. 2 is the anisotropy of the spin displacement with respect to the orientation of the wire. In particular, the nature of the anisotropy appears to depend strongly on the electron density, as well as the relative strength of the Rashba and Dresselhaus spin-orbit couplings. In the 2D limit, we saw that the transverse spin current was maximized for electron momentum parallel to the [110] axis and was minimized along the [$\bar{1}$10] axis. However, in the quasi-1D situation shown in Fig. 2, this is rarely the case. In fact, for electron densities below the second subband, the anisotropy is actually *opposite* that of the 2D case, with $\Delta x_s$ maximized when the wire is oriented along [$\bar{1}$10]. For larger electron densities, the interference of multiple subbands leads to a much more complicated pattern of anisotropy, and in many cases $\Delta x_s$ is largest along neither [110] nor [$\bar{1}$10]. Another interesting feature seen in Figs. 2a-2e is that over a small change in electron density, $\Delta x_s$ can transition from having its maximum value along the [$\bar{1}$10] axis to having its minimum value along this axis. For example, in Fig. 2b this occurs between densities of 0.8 and 0.9 x $10^{11}$ cm$^{-2}$. In this density range, the value of $\Delta x_s$ also goes from positive to negative.

The anisotropic behavior described above can be explained with a detailed look at the energies and spin displacements of the individual subbands in the quantum wire. In Fig. 3a, we plot the subband energies of the quantum wire as a function of the electron momentum along the length of the wire, for $\alpha_z$ = 20 meV-nm and $\beta$ = 5 meV-nm, corresponding to Fig. 2a. In order to highlight the role of the spin-orbit coupling, the kinetic energy term has been removed, $E_{norm}(k_\parallel) = E(k_\parallel) - \hbar^2 k_\parallel^2 / 2m^*$. The red solid lines are the subband energies for the wire oriented along [110], and the blue dotted lines are for [$\bar{1}$10]. The black dashed line is the Fermi energy corresponding to an electron density of 0.4 x $10^{11}$ cm$^{-2}$. The behavior of the subband energies echoes what has been shown in earlier works that considered spin-orbit coupling in quantum wires [31,33,34]. For low values of $k_\parallel$, there is a linear energy splitting of each subband, which is greater for [$\bar{1}$10] than it is for [110]. In Fig. 3b, we plot the spin separation of the lowest set of subbands. We see that for low values of $k_\parallel$, and for subband **1-**, the magnitude of $\Delta x_s$ is larger for [110] than for [$\bar{1}$10]. These results reflect the predictions for the 2D case in Fig. 1. For larger values of $k_\parallel$, there is an anticrossing between adjacent subbands. As shown in Fig. 3b, the anticrossing of subband **1+** is accompanied by an enhancement of its spin separation. For [$\bar{1}$10], the energy difference between the anticrossing subbands is smaller than that for [110], and the corresponding spin separation is larger. This is consistent with the topological picture discussed earlier, where the Berry curvature, and thus the strength of the spin Hall effect, is inversely proportional to the energy difference between the adjacent subbands. In Fig. 3, the solid circles indicate the points where the Fermi energy crosses the subbands for $[\bar{1}10]$, while the open squares indicate these points for $[110]$. At this density, the Fermi energy intersects subband **1+** near the anticrossing point, and therefore the overall spin separation is larger for $[\bar{1}10]$ than it is for $[110]$. This, as well as the smaller negative contribution due to subband **1-**, explains why the anisotropy of the spin Hall effect in the quantum wire at low densities is opposite that of the infinite 2D system.

In Fig. 2b, as the density increased from 0.8 to 0.9 x $10^{11}$ cm$^{-2}$, the spin separation evolved from being a maximum along the $[\bar{1}10]$ axis to being a minimum along this axis. This behavior can be explained with an examination of Fig. 4, where the energies and spin separations of each subband are plotted as a function of $k$, for $\alpha = 10$ meV-nm and a wire orientation of $[\bar{1}10]$. In Fig. 4a, the solid lines show the normalized subband energies, and the dashed lines show the Fermi energies corresponding to electron densities of 0.8 and 0.9 x $10^{11}$ cm$^{-2}$. Fig. 4b shows the corresponding spin separation of each subband. The solid circles indicate the points where subbands **1+** and **2-** cross the Fermi energy at $n_{2D} = 0.8$ x $10^{11}$ cm$^{-2}$, while the open squares show these points for $n_{2D} = 0.9$ x $10^{11}$ cm$^{-2}$. At the lower electron density, subband **1+** crosses the Fermi energy at the point where its spin separation is a maximum, which results in a positive spin separation. At the higher electron density, subband **1+** crosses the Fermi energy away from the maximum of its spin separation, while subband **2-** crosses the Fermi energy at the point where its spin separation is a minimum. Thus, in this case the overall spin separation takes on a slightly negative value. This behavior can also be seen in Figs. 2c-2e, where the value of $\alpha$ is much closer to the value of $\beta$. In these cases, the anticrossing between adjacent subbands occurs over a much smaller range of $k$, which means that along the [$\bar{1}10$] axis the transition from positive to negative spin separation occurs over a much smaller range of electron density. Finally, we note that in Fig. 2f, when $\alpha = 1$ meV-nm, the spin separation is always maximized for a wire orientation of [$\bar{1}10$]. In this situation, the small values of $\alpha$ and $\beta$ result in a small energy splitting, and the subbands in the wire never reach their anticrossing point for the range of electron densities that we considered. As discussed above, prior to the first subband anticrossing the spin separation is maximized for a wire orientation of [$\bar{1}10$].

In summary, we have used numerical simulations to investigate the intrinsic spin Hall effect in quantum wires in the presence of both Rashba and Dresselhaus spin-orbit coupling. Our results include some expected behavior and some behavior that is less obvious. Among the expected results, we saw that the overall spin separation increases with the spin-orbit coupling strength and decreases with the electron density, with the latter effect due to the interference of multiple subbands in the quantum wire. We also saw that when the Rashba spin-orbit coupling becomes weaker than the Dresselhaus spin-orbit coupling, the spin separation reverses direction, which is in line with earlier work on this phenomenon. Among our less obvious results is the nature of the anisotropy of the spin Hall effect. While anisotropy is expected in 2D systems with both Rashba and Dresselhaus spin-orbit coupling, we found that the situation is greatly complicated by the presence of lateral confinement in a quantum wire. In particular, at low densities when only one subband of the quantum wire is occupied, the spin Hall effect is strongest for electron momentum along the [$\bar{1}10$] axis, which is opposite than what is expected in the purely 2D case. In addition, when more than one subband is occupied, the strength and anisotropy of the spin Hall effect can vary greatly over relatively small changes in electron density, making it difficult to predict which wire orientation will maximize the strength of the spin Hall effect.

These results can be used to guide experimental work that utilizes the spin Hall effect in quantum wire spintronics. In particular, in order to maximize the strength of the spin Hall effect and avoid unpredictable anisotropic behavior, only the lowest subband in the wire should be occupied. This can be accomplished with an appropriate choice of wire width prior to fabrication [31], or by adjusting the density in the wire after fabrication with either a global gate or a quantum point contact that filters out the higher subbands. In either case, our results indicate that the wire should be oriented along the $[\bar{1}10]$ axis of the host semiconductor in order to take advantage of the anisotropy of the system when both Rashba and Dresselhaus spin-orbit coupling are present.

AWC acknowledges the support of the DOE Computational Science Graduate Fellowship; grant number DE-FG02-97ER25308. Special thanks also go to Ethan Coxsey for helping to run simulations.

This document is the unedited author's version of a work that has been published in *J. Phys.: Condens. Matter* **23**, 465301 (2011). To access the final edited and published work see dx.doi.org/10.1088/0953-8984/23/46/465301.

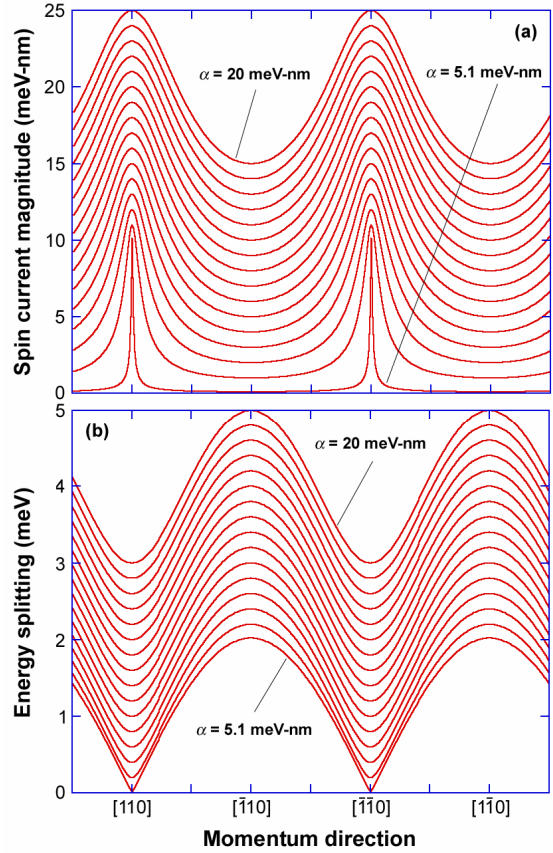

Fig. 1. Part (a) show the spin Hall current magnitude in a 2DEG, and part (b) shows the energy splitting of the conduction band, both as a function of the direction of electron momentum. In both cases, $\beta = 5$ meV-nm and $\alpha = 5.1, 6, 7, 8, \ldots, 20$ meV-nm.

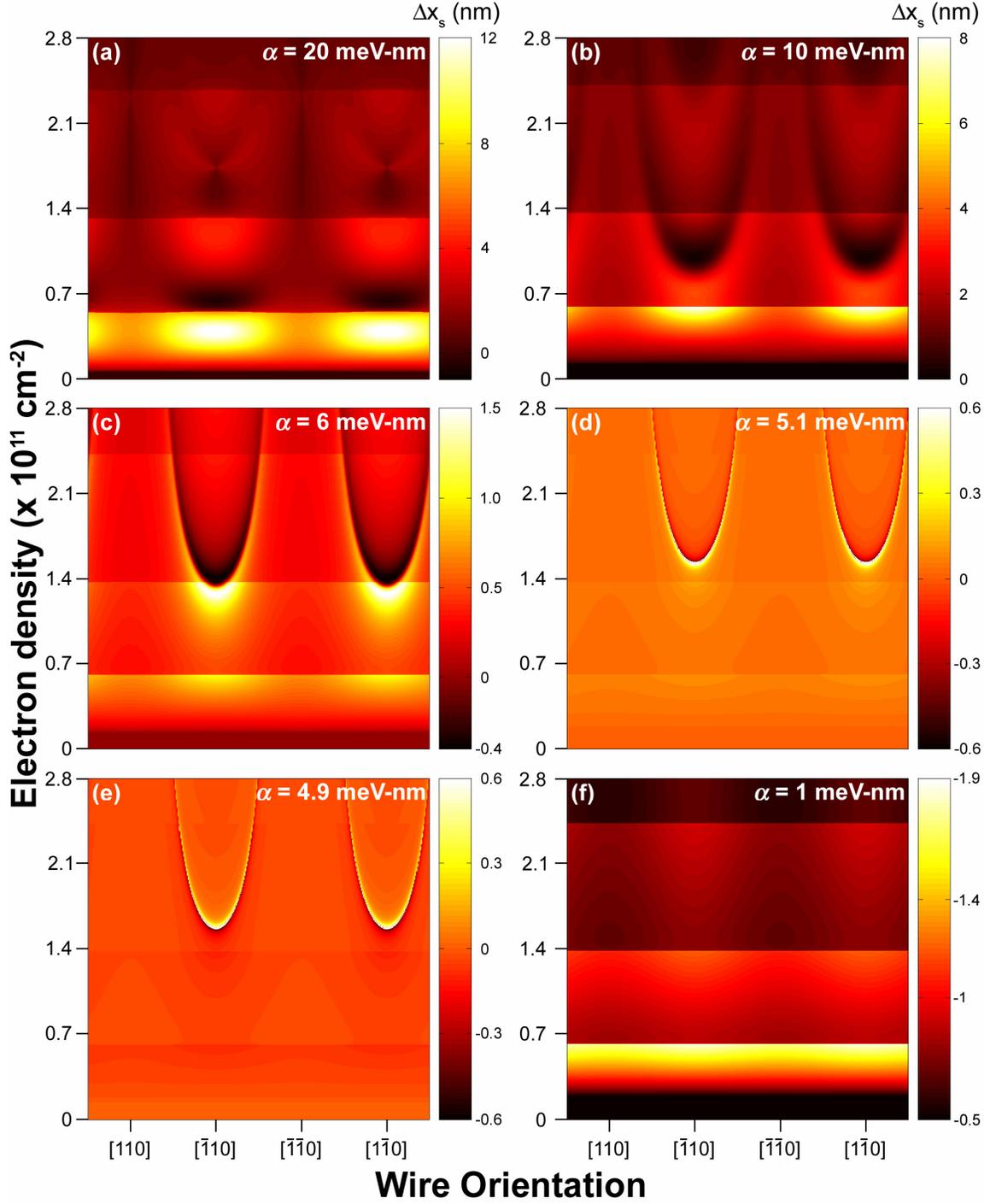

Fig. 2. Spin separation as a function of the electron density and the orientation of the quantum wire, for various values of $\alpha$. In each case, we assume a 100-nm-wide GaAs quantum wire with $\beta = 5$ meV-nm.

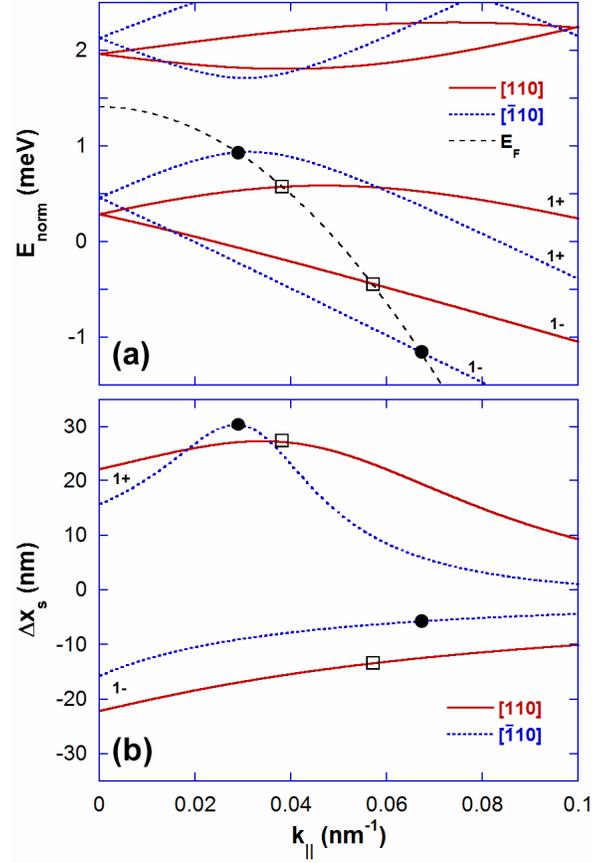

Fig. 3. Part (a) shows the subbands of the quantum wire as a function of the electron momentum along the wire, with the kinetic energy portion removed. The dashed line shows the Fermi energy corresponding to an electron density of $n_{2D} = 0.4 \times 10^{11}$ cm$^{-2}$. Part (b) shows the spin separation of the wave function associated with each subband. The solid circles show where the subbands cross the Fermi energy for a wire orientation of $[\bar{1}10]$, while the open squares show this crossing for $[110]$. In this plot, we assume a 100-nm GaAs wire with $\alpha = 20$ meV-nm and $\beta = 5$ meV-nm.

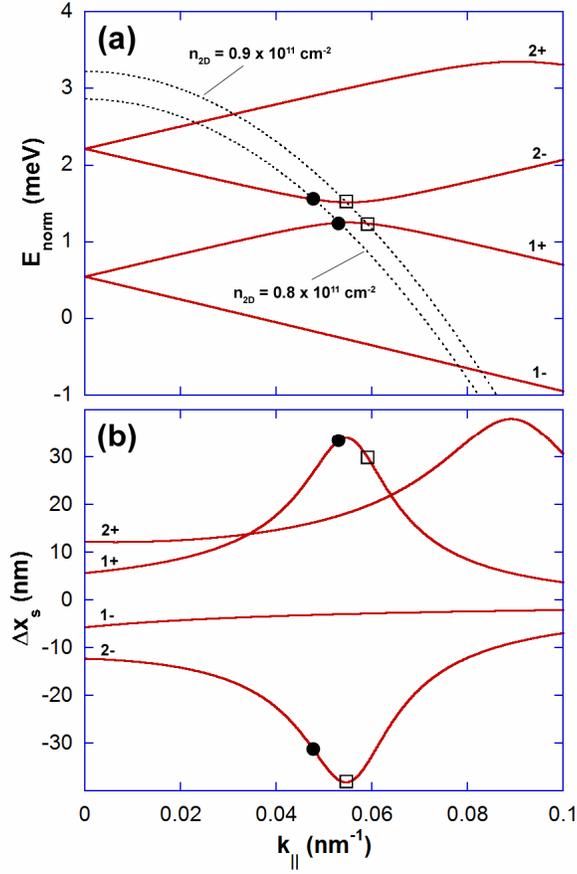

Fig. 4. Part (a) shows the subbands of the quantum wire as a function of the electron momentum along the wire, with the kinetic energy portion removed. The dashed lines indicate the Fermi energy at two different electron densities. Part (b) shows the spin separation of the wave function associated with each subband. The solid circles show where the subbands cross the Fermi energy for $n_{2D} = 0.8 \times 10^{11}$ cm$^{-2}$, while the open squares show this crossing for $n_{2D} = 0.9 \times 10^{11}$ cm$^{-2}$. We assume a 100-nm GaAs wire oriented along $[\bar{1}10]$, with $\alpha = 10$ meV-nm and $\beta = 5$ meV-nm.